\documentclass[aps,showpacs]{revtex4}
\usepackage[dvips]{graphicx}
\usepackage{amssymb}
\usepackage{amsmath}
\usepackage{amsmath, amsthm, amssymb, mathrsfs}
\usepackage[utf8]{inputenc}

\newcommand{\be}{\begin{equation}}
\newcommand{\ee}{\end{equation}}
\newcommand{\ben}{\begin{eqnarray}}
\newcommand{\een}{\end{eqnarray}}
\newcommand{\bes}{\begin{subequations}}
\newcommand{\ees}{\end{subequations}}
\newcommand{\pa}{\partial}
\newcommand{\bb}{\bibitem}

\begin{document}
\title{Perturbative aspects and conformal solutions of $F(R)$ gravity}
\author{D. Bazeia,$^1$ B. Carneiro da Cunha,$^2$ R. Menezes,$^1$ and A.Yu. Petrov$\,^1$}
\affiliation{{$^1$Departamento de F\'{\i}sica, Universidade Federal da
Para\'{\i}ba, 58051-970, Jo\~ao Pessoa PB, Brazil}
\\{$^2$Departamento de Física, Universidade Federal de
Pernambuco, 50670-901 Recife PE, Brazil}}

\begin{abstract}
We investigate perturbative aspects of gravity with a general $F(R)$
Lagrangian, as well as nonperturbative dilatonic solutions. For the
first part, we are interested in stability and the definition of
asymptotic charges. The main result of this study is that, while
generic $F(R)$ theories are stable under metric perturbations, they
are expected to show instabilities against curvature perturbations
when the Lagrangian includes $1/R$ terms. For the second part, one
is interested on exact solutions, and we explicitly construct
kink-like solutions of the Liouville type for the dilaton field
for $F(R)$ having the form $R+\gamma R^n,$ in two and in four dimensions.
\end{abstract}


\maketitle

\section{Introduction}

In this work we investigate perturbative and nonperturbative effects
in gravity, motivated by the cosmological constant problem
\cite{we}. As one knows, the cosmological constant engenders an
important but very hard problem to solve due to both cosmological and
phenomenological reasons. In the cosmic evolution of the Universe,
particle physics introduces the main mechanisms to govern the several
phases of the Universe, and cosmology drives the way the cosmological
constant is accounted for in each phase. However, the presence of
spontaneous symmetry breaking, which is the basic mechanism underlying
unification of the fundamental interactions in nature, inevitably
injects vacuum energy density even if no such term is present in the
theory. For this reason, one realizes that not only the physics at the
Planck scale, but also the low energy infrared dynamics of gravity
itself is essential to the resolution of the cosmological constant
problem. The reasoning establishes direct connection between
nonperturbative effects in gravity and the cosmological constant
problem.

Nowadays, the cosmological constant problem must include the important discovery that the Universe
is currently evolving in an accelerating phase \cite{acce}. There are several distinct possibilities of taking
acceleration into account: from one side, we can keep standard
geometry, incorporating modifications on the matter contents of the
model. Interesting investigations which deal with this possibility
include, for instance, the cosmological constant \cite{we}, dynamical
scalar fields \cite{rp}, Chaplygin fluid \cite{pa}, and phantom
dynamics \cite{cal}. Another possibility deals with modifications of
geometry, changing General Relativity; here an important class of
theories includes models which depend on the Ricci scalar, usualy
named $F(R)$ theories \cite{fr}. In the present work, we will deal
with the latter, that is, we will investigate extended
gravity, changing $R\to F(R),$ for $F(R)=R+\gamma R^n,$ including the
cases $n=2$ and $n=-1.$ This modification brings interesting
novelties, as we will show below. The motivation goes beyond the
recent discovery of the accelerated expansion of the Universe
\cite{acce}, and it is very natural to expect that, due to the fact
that the curvature of the present Universe is very small, the
nonperturbative effects should be characteristic for some typical
forms of $F(R).$ In particular, the interest in $n=2$ also appear from
semi-classical investigations, and the case of $F(R)=\exp(-\alpha R)$
suggests the presence of instanton effects.

The investigation of nonperturbative solutions is of direct interest
to a diversity of applications, including black hole, integrability,
noncommutative geometry and other problems \cite{di,stro}. The
nonperturbative issues are in
general hard to solve, mainly in high energy physics in curved
space-time. For this reason, we are planning to approach the problem
with the use of the nonperturbative procedure referred to as the
first-order formalism, in which one solves equations of motion with
solutions of first-order differential equations. Some of us have
recently done some progress in extending the formalism to curved
space-time \cite{fof}, and we now consider this possibility in
connection with dilaton gravity.

In the present study we organize this Letter as follows. In the next
Sec.~\ref{sec:2} we derive the equations of motion for $F(R)$
gravity theories and the conserved charges, and then we study stability of solutions
in these theories. In Sec.~\ref{sec:3} we turn attention to gravity in the conformal
sector, and there we describe the $F(R)$ model restricted to the conformal metric,
investigating the corresponding equations of motion in two and in four dimensions.
We end the work in Sec.~{\ref{sec:end}}, where we include some final considerations.

\section{Equations of motion and conserved charges}
\label{sec:2}

Let us lay down the notation, starting with the $F(R)$ action
\be
\label{fr}
S=-\int d^Dx\sqrt{|g|}F(R)
\ee 
where $F(R)$ is equal to $R+2\Lambda$ for the usual Einstein-Hilbert
action with a cosmological constant $\Lambda$. The equations of motion for this theory look like
\be
\frac{1}{2}g_{ac}F(R)-F'(R)R_{ac} - F''(R)[\nabla_a\nabla_c R  - \nabla_b\nabla^b R
g_{ac}] - F'''(R)[\nabla_a R \nabla_c R - g_{ac}\nabla_d R\nabla^d R] =0 \label{eom}
\ee

Let us now compute the conserved charge associated with this
Lagrangian, following the general procedure of \cite{Iyer:1994ys}. From above, the boundary term one obtains from the
Lagrangian is
\be
\theta^b=-F'(R)(\nabla_a(\delta g^{ab})-\nabla^b(\delta g))-\nabla_a
F'(R)\,\delta g^{ab}+\nabla^b F'(R)(\delta g)
\ee
that is to say, the full variation of the action can be written as
$\delta \mathscr{L}=\sqrt{|g|}E_{ab}\,\delta g^{ab}+\sqrt{|g|}\nabla_a \theta^a,$
where $E_{ab}$ are the equations of motion (\ref{eom}). Now, the
Noether procedure gives us a conserved current associated with a
symmetry generated by a vector field $\xi^a$. Those types of
diffeomorphisms change the metric as $\delta g_{ab}=\nabla_a\xi_b+\nabla_b\xi_a$.
The conserved current is $J^a[\xi^b]=\theta^a-\xi^a\mathscr{L}.\label{current1}.$
The existence of this current stems directly from the proof of the
theorem. The hard step in any theory which is invariant under 
diffeomorphisms is to find the charge associated with $J^a$. Consider
the Poincar\'e dual to $J^a$: ${\bf J}_{a_1\ldots a_{n-1}}=\epsilon_{aa_1\ldots a_{n-1}}J^a,$
where $\epsilon_{a_1\ldots a_n}$ is the volume form associated with
the metric. We will exceptionally use $n$ for the dimension of
space-time in this subsection to avoid cluttering the
notation. Conservation of the current implies that, {\it on shell}, 
the $n-1$ form ${\bf J}$ is closed $d{\bf J}=0$, so it can be written
locally as a exterior derivative: ${\bf J}_{a_1\ldots a_{n-1}}=\nabla_{[a_1}{\bf Q}_{a_2\ldots a_{n-1}]}
=\partial_{[a_1}{\bf Q}_{a_2\ldots a_{n-1}]},$ or, analogously with the duals
$J^b=\nabla_a Q^{[ab]}.$ We can check directly that this equation above implies the
conservation of the current
\be
\nabla_b J^b=\nabla_b\nabla_a Q^{[ab]}= -\frac{1}{2}[\nabla_a,
\nabla_b]Q^{[ab]}=\frac{1}{2}({R_{abc}}^aQ^{[cb]}+{R_{abc}}^b
Q^{[ac]})=\frac{1}{2}(-R_{bc}Q^{[cb]}+R_{ac}Q^{[ac]})=R_{ac}Q^{[ac]}=0
\ee

If we are able to find such $Q^{ab}$ we can write directly the
conserved charge by integrating the corresponding $n-2$ form ${\bf Q}$
over a suitable $n-2$ hypersurface. The procedure is then analogous to
the definition of conserved charges in field theory which can be
written under certain assumptions as the integral of a charge flux
over a sphere (a $n-2$ dimensional hypersurface in $\mathbb{R}^{3,1}$)
``at infinity''. 

As it turns out, one can indeed write a $Q^{ab}$ for the theory
above; we suppose it has the form $Q^{ab}[\xi^c]=2W^{[a}\xi^{b]}+2X\nabla^{[a}\xi^{b]},$
for some metric-dependent $W^a$ and $X$. Taking its divergence, we will find several terms 
\be
\nabla_a Q^{ab}= \xi^b\nabla_aW^a-\xi^a\nabla_a W^b+ W^a\nabla_a\xi^b
-W^b\nabla_a\xi^a+\nabla_a X (\nabla^a\xi^b-\nabla^b\xi^a)+ X(\nabla_a
\nabla^a\xi^b-\nabla_a\nabla^b\xi^a) 
\label{chargediv}
\ee
On the other hand, the direct calculation from (\ref{current1}) gives
\be
J^b= F'(R)[\nabla_a(\nabla^a\xi^b+\nabla^b\xi^a)-
2\nabla^b\nabla_a\xi^a]-F''(R)\nabla_a R(\nabla^a\xi^b+\nabla^b\xi^a
-2g^{ab}\nabla_c\xi^c)
-2 F'(R)\nabla^b R \nabla_a\xi^a+\xi^b F(R)
\label{current2}
\ee
Comparing the terms involving two derivatives of $\xi^a$, we find that
$X$ must be $X=F'(R),$ up to some term proportional to the equations of motion.
The terms involving only one derivative of $\xi^a$
\be
W^a\nabla_a\xi^b-W^b\nabla_a\xi^a + \nabla_a X (\nabla^a\xi^b -
\nabla^b\xi^a)\stackrel{?}{=} -F''(R)\nabla_a R[\nabla^a\xi^b
+\nabla^b\xi^a-2g^{ab}\nabla_c\xi^c]
\ee
which gives $W_b=-2F''(R)\nabla_bR.$ We now compare the
term proportional to $\xi^a$ in (\ref{chargediv}) and (\ref{current2}) to get
\be
\xi^b\nabla_a W^a-\xi^c\nabla_c
W^b\stackrel{?}{=}-2F'(R){R^b}_d\xi^d+F(R)\xi^d
=(F(R)g^{ab}-2F'(R)R^{ab})\xi_a 
\label{linearorder}
\ee
With the choices above for $X$ and $W_b$ one has, for the left-hand side
\be
\begin{aligned}
\xi^b\nabla_a W^a-\xi^c\nabla_c W^b &= -2 F'''(R)\xi^b\nabla_a R
\nabla^aR -2F''(R)\xi^b\nabla^2 R + 2 F'''(R)\xi^a\nabla_aR \nabla^bR
+2F''(R)\xi^a\nabla_a\nabla^bR  \\ 
&=[2F'''(R)(\nabla^aR\nabla^bR-g^{ab}\nabla_cR\nabla^cR)+2F''(R)
(\nabla^a\nabla^b R-g^{ab}\nabla_c\nabla^cR)]\xi_a
\end{aligned}
\ee
which is the left-hand side of the equation (\ref{linearorder}) by the
equations of motion (\ref{eom}). The conserved charge is thus given by
$Q^{ab}=2F'(R)\nabla^{[a}\xi^{b]}+4F''(R)\xi^{[a}\nabla^{b]}R.$
In this expression the first term is analogous to the Komar mass
term in usual Einstein-Hilbert whereas the second term is new.

Now we apply the expression for the charge found above
to spherically symmetric spaces. As all of the solutions discussed
here are static, the total mass will be given by the expression above
when the vector field associated is the time translation vector
$\xi^a=(\partial_t)^a$, which is hypersurface orthogonal and its
derivative is given by
$\nabla_a\xi_b=-\xi_{[a}\nabla_{b]}\log|\xi_c\xi^c|=(dt)_{[a}
(dr)_{b]}\,d_rf(r) =\epsilon_{ab}f'(r),$
where in the first step we used the general expression found in
equation (C.3.12) from \cite{Wald}, the second step uses the fact that
$g_{ab}\xi^b=f(r) (dt)_a$, and the third step defines the volume
element of the ``$t-r$'' plane as $\epsilon_{ab}=
(dt)_{[a}(dr)_{b]}$. Last but not least, we have used the chain rule
to expand $\nabla_af(r)=-(dx)_af'(r)$. One can also derive
this result by noting that the relevant part of the metric for
spherically symmetric spaces is given by
$ds^2_{r-t}=f(r)(dt)^2-(dr)^2/f(r)$. Notice that the
expression for the derivative of $\xi^a$ above allows us to write the
charge as a term proportional to the spatial volume element
$\xi^a\epsilon_{aa_1\cdots a_{n-1}}$:
${\bf Q}_{a_1\cdots a_{n-2}}=-2(F'(R)\nabla^b\log|\xi_c\xi^c|-
2F''(R)\nabla^bR)\xi^a\epsilon_{aba_1\cdots a_{n-2}}.$
Applying the same procedure for $\nabla_aR$, we can write the charge
as 
\be
Q_{ab}=2(F'(R)\,f'(r)-2F''(R)\,R')\epsilon_{ab}.
\ee
We note that, since the space is spherically symmetric, the
integral over a $D-2$ dimensional sphere one has to perform to compute
the mass is trivial. One should remark that this expression is really
better suited for asymptotically flat space-times. The definition of
mass on non-asymptotically flat spaces is murky, where one usually
uses the trick of ``shifting'' the value of the mass by subtracting
the ``vacuum'' value, {\it i. e.}, the value of ${\bf Q}$ on a
maximally symmetric solution with the same asymptotics. We would to
conclude by commenting that flat-space-solutions like Schwarzschild-de
Sitter will have its mass modified by the presence of the $F'(R)$
term, although not by much. One does expect, however, that solutions
with long-range modifications of the curvature, such as those with
$R\approx 1/r$ asymptotically, will show a dramatic effect, specially
for $F(R)$ involving $1/R$ terms, for then $F''(R)$ can be quite
large. 

Let us now focus attention on stability. We investigate small deviations of an Einsteinian space
solution $R_{ab}=\frac{R}{D} g_{ab}$ of the general equation of motion (\ref{eom}). In order to do this, let 
us take the variations of the Ricci tensor and the scalar curvature
and write them in terms of the linearized metric components
$h_{ab}=\delta g_{ab}$: 
\be
\delta R_{ab} = \frac{1}{2}\nabla^2 h_{ac}
+\frac{1}{2}\nabla_a\nabla_c h -\nabla_b\nabla_{(a}{h_{c)}}^b 
 = \frac{1}{2}\nabla^2
\bar{h}_{ac} -\frac{1}{2(D-2)}\nabla^2\bar{h}g_{ac} -\nabla_b\nabla_{(a}\mbox{$\bar{h}_{c)}$}^b
\label{variationricci1} 
\ee
and hence
\be
\delta R = -h^{ab}R_{ab}+g^{ac}\delta R_{ac}  =
\frac{2R}{D(D-2)}\bar{h}-\frac{1}{D-2}\nabla^2 
\bar{h}. \label{variationscalar} 
\ee
In the expressions above,
$\bar{h}_{ab}=h_{ab}-\frac{1}{2}hg_{ab}$. Such a choice stems from the
gauge invariance of linearized gravity: $h_{ab}$ and
$h_{ab}+\nabla_av_b+\nabla_bv_a$ correspond to the same
perturbation \cite{note}. By writing the variations in this form one can get
rid of the first term in (\ref{variationricci1}) by choosing $v_a$ such
that: $\nabla^2v_a+{R_a}^bv_b=-\nabla_{b}\mbox{$\bar{h}_{a}$}^b.$
This fixes some of the gauge symmetry, but the equation above
determines $v_a$ up to a vector $w_a$ which satisfies $\nabla^2w_a+{R_a}^bw_b=0.$
We should point out that solving the two equations above amount to
find a vector potencial in Lorentz gauge for some distribution of
current. With this choice, we can write the first term in
(\ref{variationricci1}) as
\be
\begin{aligned}
\nabla_b\nabla_{(a}\mbox{$\bar{h}_{c)}$}^b &= \nabla_{(a}\nabla_{|b|} 
\mbox{$\bar{h}_{c)}$}^b +
[\nabla_b,\nabla_{(a}]\mbox{$\bar{h}_{c)}$}^b 
=-R^d_{(ca)b}\mbox{$\bar{h}_{d}$}^b+
R^b_{db(a}\mbox{$\bar{h}_{c)}$}^d= 
-{R^d_{cab}}\mbox{$\bar{h}_{d}$}^b +R_{d(a}\mbox{$\bar{h}_{c)}$}^d
\end{aligned}
\ee

We can now turn to the variation of the equations of motion
(\ref{eom}). The relevant terms are
\be
\frac{1}{2}h_{ac}F(R)+\frac{1}{2}g_{ac}F'(R)\delta R - F''(R)\delta R
R_{ac} -F'(R)\delta R_{ac}-F''(R)[\nabla_a\nabla_c \delta R -\nabla^2
\delta R g_{ac} ] =0 \label{2ndvariation}
\ee
and we omitted terms which involve derivatives of $R$, which are zero
on shell since $R$ is assumed constant. One can further
simplify the equation above by noting that the variation of the
curvature scalar only depends on $\bar{h}$, or, equivalently $h$, the
trace of the metric variation. By taking the trace of the equation
above, on the absence of matter, one finds a fourth order equation for
$\bar{h}$, which can yield the trivial solution $h=0$ if one picks the
right initial values for the lower derivatives of $h$. This can be
accomplished by a suitable gauge transformation for $w_a$. So, in
short, we can, in a general $F(R)$ theory, accomplish the transverse
traceless gauge for metric perturbations: $\nabla_ah^{ab}=0,$ $h=0.$
In this case $\bar{h}_{ab}=h_{ab}$ and then the only relevant terms in
(\ref{2ndvariation}) are those that depend on $h_{ab}$. Picking those,
we have a simple equation for the metric perturbations
\be
\frac{1}{2}F(R)h_{ac}-
F'(R)\left(\frac{1}{2}\nabla^2h_{ac}-{{R^b}_{ac}}^{\raisebox{-0.9mm}{$\scriptstyle d$}}  h_{bd}
+R_{d(a}{h_{c)}}^d\right)=0
\label{perturbations}
\ee
or, specializing to Einstein spaces,
\be
\nabla^2h_{ac}+2{{C^b}_{ac}}^{\raisebox{-0.9mm}{$\scriptstyle d$}}
h_{bd} 
+\frac{2R}{D-1}h_{ac}-\frac{F(R)}{F'(R)}h_{ac}=0,
\ee
where $C_{abcd}$ is the Weyl tensor. The causal structure of the
equation above for Einstein spaces has been studied in
\cite{Aragone:1971kh}, and we will take the result of a particularly
nice survey of \cite{Buchbinder:2000fy}. As one can see from, for
instance, the behavior of the equation above under conformal
transformations, the causal structure of the equation above is
governed by a ``mass term'' which depends on the spin of the particle
as well as the scalar curvature. In this case, the effect is that the
mass term $M^2$ of spin 2 fields is modified by the presence of
curvature to
\be
M^2=-\frac{F(R)}{F'(R)}+\frac{2}{D}R \label{masspert}
\ee
Note that this means that the term proportional to the Riemann tensor
in (\ref{perturbations}) takes into account the causal structure of
spin-2 particles and hence only the following terms contribute to the
mass of the perturbations. This is essentially the spin-2 version of
the Breitenlohner-Freedman found \cite{Breitenlohner:1982jf} in the
study of anti-de Sitter spaces. One can also check that constant
curvature solutions of the generic Lagrangian (which includes
Einstein-Hilbert with a cosmological constant as a particular case)
always have massless perturbations (gravitons). Therefore, these
solutions do not present us with new instabilities. These are
essentially the same which arise in the study of Einstein-Hilbert
Lagrangian.

The equation above (\ref{masspert}) gives a straightforward criterion
for the stability of the solutions described below: if the effective
squared mass of the gravitons is positive, small deviations of the
metrics found will oscillate, otherwise they will grow exponentially
and the space found will be unstable. For the special case $D=4$, the
corresponding equation has been found by several authors -- see \cite{bunch}
and references therein.

Theories involving $1/R$ terms, which we call
``non-perturbative,'' has attracted some attention recently, mainly
because of the prospect of having a solution of positive curvature,
{\it i.e.}, a assymptotically de Sitter space-time, without the need
for a cosmological constant. We analyze one
well-known solution of those theories which shows an instability against
curvature perturbations which should plague those theories in general.

We remark that the mere existence of a term
proportional to the inverse of the curvature scalar does not fit with
what we know from renormalizable field theory, or even what little we
can grasp of quantum gravity, as in string theories. Usual corrections
from perturbation series in the interactions of quantum fields with
classical gravity show the dynamical generation of terms up to
$R^{D/2}$ \cite{Birrell}. In string theory, general remarks given in
\cite{Seiberg1986} can be applied here to constrain what corrections
one can expect from string instantons such as those that arise from
compactifications which respect $N=1$ SUGRA. Such compactifications
have received a lot of attention in recent years since the work of
\cite{Kachru:2003aw} introduced extra ingredients like confined
fluxes, which allowed for meta-stable de Sitter vacua. At any rate,
instantons and fluxes may break the Super-Weyl rescaling symmetry of
SUGRA [$g_{ab}\longrightarrow e^{2\lambda}g_{ab},$ and $
e^{2\lambda}=-3/{(\Phi^+_i\Phi_i+c_i\Phi_i+\bar{c}_i\Phi^+_i-3),}$
where $\Phi_i$ is a chiral superfield] to a discrete set, as it
happens with other superfields. The residual symmetry constrain the
effective potential generated to be of the form $e^{-\alpha R}$. In
this expression, $\alpha$ incorporates scalar condensates, but should
not depend on the metric. One should note that even in this
hypotetical scenario, the corrections one expects from quantum strings
are perturbative in $R$.

All the preceding discussion may present an obstacle for the
consideration of non-perturbative theories, and below we will present
a concrete criticism, following the study of \cite{Dolgov:2003px},
while keeping the dimension arbitrary. We consider the general
equation of motion (\ref{eom}) for 
$F(R)=R-\gamma^2 /R$. For $R=R_0$ constant, the trace of the equation
gives an algebraic condition:
\be
\frac{D}{2}\left(R-\frac{\gamma^2}{R}\right)=
\left(1+\frac{\gamma^2}{R^2}\right)R
~~\Longrightarrow~~
R_0^2=\frac{D+2}{D-2}\gamma^2
\ee
which gives a small curvature for $\gamma$ small, as desired. We will
be interested on the positive curvature solutions. Now
consider a metric perturbation such that the curvature scalar is no
longer a constant. Taking the trace of (\ref{eom}) we find that 
\be
-(D-1)(F''\nabla^2R+F'''(\nabla R)^2)=
-(D-1)\nabla^2(F'(R))=\frac{D}{2} F(R)- F'(R) R. 
\ee
Taking a small curvature perturbation $R=R_0+\delta R$, we will have
\be
(D-1)F''(R_0)\nabla^2\,\delta R\approx-\left(\frac{D-2}{2}F'(R_0)-
  F''(R_0)R_0\right)\delta R
\ee
where we omitted a term involving one derivative of $\delta R$, which
depends on the particular metric perturbation but will not change the
conclusion. The equation above gives an effective mass for such
perturbations 
\be
m^2\approx -\frac{1}{D-1}R_0+\frac{D-2}{2(D-1)}
\frac{F'(R_0)}{F''(R_0)} \label{effectivemass}
\ee
Note that, if we have the Einstein-Hilbert term, $F'(R_0)\approx 1 +
{\cal O}(R_0)$. Now, for perturbative theories, $F''(R_0)\approx 1$ is
independent of $R$ and then the last term of the right-hand side
dominates for small curvatures. However, for non-perturbative
theories like $F(R)=R-\gamma^2/R$ we will have problems; indeed, in
this case we get 
\be
m^2\approx -\frac{D+2}{2(D-1)}R_0,
\ee
which shows that such theories are generically unstable under
curvature perturbations. We note that this result is below the
negative mass bound since $R_0>0$. One may wonder, like in
\cite{Nojiri:2003ni}, whether higher order terms, perturbative in $R$,
may help the situation. From the discussion above, one sees that if a
term like $-\gamma^2/R$ is present, then $F''(R)= -2\gamma^2/R^3+{\cal O}(R)$.
For the (small) constant curvature solution, $R\propto
\gamma$, and then $F''(R)$ is dominated by the non-perturbative term,
while the first derivative of $F(R)$ is still of order $\gamma^0$,
hence the expression (\ref{effectivemass}) should not change
considerably as higher order terms in $R$ are added to the
Lagrangian. One should also keep in mind that the same arguments given
above against the naturalness of a non-perturbative term to the
Lagrangian also point to the fact that the numerical coefficients in
from of higher order terms are too small to have any dramatic effect
in the dicussion above, as can be seen in \cite{Birrell}.

\section{The dilatonic sector}
\label{sec:3}

Let us now study the conformally flat space
\cite{di}, where the dynamics is restricted to the conformal
sector and the metric tensor $g_{\mu\nu}(x)$ is given by $g_{\mu\nu}(x)=e^{\sigma(x)}\eta_{\mu\nu},$
with $\sigma(x)$ standing for the conformal factor, the dilaton field.
In this case, the scalar curvature has the form 
\be
\label{curv}
R=g^{\alpha\beta}R_{\alpha\beta}=e^{-\sigma}\left(\frac{D-1}{2}\right)\Big[2\Box\sigma+\left(\frac{D-2}{2}\right)\pa^{\alpha}\sigma\pa_{\alpha}\sigma\Big].
 \ee
The standard Einstein-Hilbert action which includes the cosmological
constant is given by $S=-\int d^Dx\sqrt{|g|}(R+2\Lambda),$ where we are using $8\pi G=1.$
The action modificated by the change $R\to F(R)$ has now the form
(\ref{fr}). More explicitly, we can write
 \be
S=-\int d^Dxe^{\frac{D}{2}\sigma}F\left(e^{-\sigma}\left(\frac{D-1}{2}\right)\Big[2\Box\sigma+\left(\frac{D-2}{2}\right)
\pa^{\alpha}\sigma\pa_{\alpha}\sigma\Big]\right)
 \ee 
This form shows cleary that for $D=2$, and $F(R)=R+2\Lambda$, the
action reduces to total derivative, whereas for $D=1$ the action is
reduced to the cosmological term alone. 

The equation of motion looks like
$De^{\frac{D}{2}\sigma}F(R)+2e^{\frac{D}{2}\sigma}
F^{\prime}(R)\frac{\delta R}{\delta\sigma}=0,$
where prime represents derivative, that is, $F^{\prime}(R)\equiv{dF}/{dR}$. 
In more manifest form, after substitution of the expression
(\ref{curv}) for the scalar curvature $R$, we have 
\ben\label{motgen}
&&D e^{\frac{D}{2}\sigma}F(R)-(D-1)F^{\prime}(R)e^{(\frac{D}{2}-1)\sigma}\Big[2\Box\sigma+
\left(\frac{D-2}{2}\right)\pa^{\alpha}\sigma\pa_{\alpha}\sigma\Big]+\nonumber\\
&+&2(D-1)\Box(e^{(\frac{D}{2}-1)\sigma}F^{\prime}(R))-{(D-1)(D-2)}\pa_{\alpha}
(e^{(\frac{D}{2}-1)\sigma}F^{\prime}(R)\pa^{\alpha}\sigma)=0.
\een

This is the equation of motion we have to deal with to find
nonperturbative solutions in dilaton gravity, modified to include the
change $R\to F(R).$ Due to intrinsic difficulties, in the following we
first investigate the simpler $D=2$ case. The strategy used in the following will then be
extended to the case $D=4,$ which is treated afterwards. 

The case $D=2$ is motivated by the well-known fact that gravity in
$(1,1)$ dimensions has the direct advantage of eliminating intricate
structures which are present in higher dimensions, easing the
investigation due to technical simplifications specific of the $(1,1)$
dimensions. From the equation of motion for $D$ arbitrary, we see that
in the $D=2$ case the equation simplifies to give 
\ben
\label{eq2}
&&e^{\sigma}F(R)-F^{\prime}(R)\Box\sigma+\Box
F^{\prime}(R)=0. 
\een
We follow the first-order formalism of Ref.~\cite{fof}, so we
concentrate on the presence of static solutions, searching for a
first-order equation connecting the derivative of the dilaton with
some specific function of it. The static dilaton only depends on
$x\equiv x^1$. Using the signature $(+,-)$ for the two-dimensional
case, we can write the scalar curvature as 
\be\label{curv1}
R=-e^{-\sigma}\sigma^{\prime\prime},
\ee
where we are using the notation
$\sigma^\prime={d\sigma}/{dx},\,\sigma^{\prime\prime}={d^2\sigma}/{dx^2},$
etc. In this case, the equation of motion (\ref{eq2}) takes the form
$e^{\sigma}F(R)+F^{\prime}(R)\sigma^{\prime\prime}-{d^2 F^{\prime}(R)}/{dx^2}=0.$
Direct calculation leads to ${d^2F^{\prime}(R)}/{dx^2}=F^{\prime\prime\prime}(R)
\cdot(\sigma^\prime\sigma^{\prime\prime}-\sigma^{\prime\prime\prime})^2 e^{-2\sigma}+
F^{\prime\prime}(R)\cdot e^{-\sigma}\left(\sigma^{\prime\prime2}+2\sigma^\prime\sigma^{\prime\prime\prime}-
\sigma^{\prime2}\sigma^{\prime\prime}-\sigma^{\prime\prime\prime\prime}\right),$
and so the equation of motion gets to the new form
\ben
\label{eqmot0}
&&e^{\sigma}F(R)+F^{\prime}(R)\sigma^{\prime\prime}-F^{\prime\prime\prime}(R)\cdot(\sigma^\prime\sigma^{\prime\prime}-
\sigma^{\prime\prime\prime})^2 e^{-2\sigma}-F^{\prime\prime}(R)\cdot e^{-\sigma}\left(\sigma^{\prime\prime2}+2\sigma^\prime\sigma^{\prime\prime\prime}-
\sigma^{\prime2}\sigma^{\prime\prime}-\sigma^{\prime\prime\prime\prime}\right)=0.
\een

We now proceed with solving this equation. Our first strategy is to
choose $F(R)$ in the form $F(R)=R+\gamma R^n+2\Lambda,$ with $\gamma$ being a constant.
Our second strategy for solving the Eq. (\ref{eqmot0}) is to choose the ansatz 
\ben
\label{ansatz}
\sigma^\prime=-2Ae^{a\sigma/2},
\een
This is the first-order equation we need: it is inspired in \cite{fof} and the specific form follows from the particular
profile of the equation of motion (\ref{eqmot0}). The first-order
equation allows writing $\sigma^{\prime\prime}=2aA^2e^{a\sigma}$,
$\sigma^{\prime\prime\prime}=-4a^2A^3e^{3a\sigma/2}$,
$\sigma^{\prime\prime\prime\prime}=12a^3A^4e^{2a\sigma}$. Therefore,
we find the following expressions: $R=-2aA^2e^{(a-1)\sigma},$ and
$\sigma^\prime\sigma^{\prime\prime}-\sigma^{\prime\prime\prime}=4A^3a(a-1)e^{3a\sigma/2},$ and
$\sigma^{\prime\prime2}+2\sigma^\prime\sigma^{\prime\prime\prime}-\sigma^{\prime2}\sigma^{\prime\prime}-\sigma^{\prime\prime\prime\prime}=-4aA^4(a-1)(3a-2)e^{2a\sigma}.$
Substituting these expressions into Eq. (\ref{eqmot0}) leads to the simpler equation 
\ben
\label{eqmotour}
e^{n(a-1)\sigma}(-2aA^2)^n (n-1)\left( 1-\frac2a n(n-2)(a-1)^2-\frac1a n(a-1)\left(3a-2\right)\right)-\frac{2\Lambda}{\gamma}=0.
\een
which can be examined in the two distinct cases $\Lambda\neq 0$ and
$\Lambda=0$. The cases $n=0$ and $n=1$ are trivial and will not be
considered. 

If {\bf $\Lambda\neq 0$}, we have to make $a=1.$ Thus, for $a=1$ the
equation (\ref{eqmotour}) reduces to the following equation for the
value of $A$: $(-2A^2)^n(n-1)-{2\Lambda}/{\gamma}=0.$
For $n$ odd we get
$A=\pm\frac{1}{2}\sqrt{-\left({2\Lambda}/{(\gamma(n-1))}\right)^{1/n}}$, and for
$n$ even, $A=\pm\frac{1}{2}\sqrt[2n]{{2\Lambda}/{(\gamma(n-1))}}$. 
These values of $A$ should be substituted to the ansatz equation
(\ref{ansatz}) which yields the solution 
\ben
\label{sol}
\sigma(x)=-2\ln(Ax+C),
\een
where $C$ is an integration constant. This is similar to the Liouville-like kink
solution found in Ref.~\cite{jac}. We note that, for
$a=1$ the curvature given above is constant and negative, equal to $R=-A^2;$ thus, we have just found an
anti de Sitter (adS) solution. 

Note also that the condition of reality of the dilaton field
$\sigma(x)$ restricts possible signs of $\Lambda$ and $\gamma,$ and
the situation can differ for different values of $n$. For example, in the case
$n=-1$ we get  
\ben
\sigma(x)=-2\ln\left(\pm\frac{1}{2}\sqrt{\frac{\gamma}{\Lambda}}x+C\right)
\een
As a result, we find that both the constants $\Lambda$ and $\gamma$ should
be either positive or negative. In the other interesting case, $n=2$,
one finds that $A=\pm\sqrt[4]{{\Lambda}/{\gamma}}$. 

If {\bf $\Lambda=0$}, we do not need to choose $a=1$ anymore; instead,
the equation (\ref{eqmotour}) is now reduced to the following equation
relating $a$ and $n$: $1-(2n/a)(n-2){(a-1)^2}- (3/a) (a-1)\left(a-2/3\right)=0.$
The solutions of this equation are $a_1={2n}/{(2n-1)},$ and $a_2={(n-1)}/{n}.$  
The ansatz equation in this case yields
\ben
\sigma(x)=-\frac{2}{a}\ln(Aax+C).\label{gensol1}
\een
Again, we get to similar Liouville-like kink solution. However, in
this case the value of $A$ cannot be fixed. We notice that the $n=1/2$
case is very specific, yielding the single root $a=-1.$ Also, we note
that this solution does not depend on $\gamma$. 

Returning to the cases $n=2$ and $n=-1$, we find that in the
$\Lambda=0$ situation both these case are possible yielding for $n=-1$
$a={2}/{3}$ and $a=2$ (for instance, for $a=2$ the solution looks like
$\sigma(x)=-\ln(2Ax+C)$) whereas for $n=2$ one finds $a_1={4}/{3}$ and
$a_2={1}/{2}$. We note that in this case, the constant curvature is
not a solution anymore. The curvature is
$R=-aA^2e^{(a-1)\sigma}=-aA^2(Aax+C)^{\frac{2}{a}-2}$, so one can see
that for $a>1$ the curvature decays to zero as $|x|$ increases to
large values. We find that for $n=2$ the localized solution gives
$a={4}/{3}$, whereas for $n=-1$ one obtains $a=2$. 

Perhaps the most interesting result is that the value $\Lambda=0$
somehow induces a phase transition: although the solutions do not
change appreciably, the scalar curvature cannot be constant in the
limit of a vanishing cosmological constant.

Before ending this section, let us examine the possibility of finding
adS solutions for $F(R)$ general. Here we keep considering the
first-order ansatz equation (\ref{ansatz}) for the dilaton, thus the
curvature (\ref{curv1}) yields 
\be
R=-2aA^2 e^{(a-1)\sigma} \label{curvature2d}
\ee
We get to constant curvature with $a=1,$ leading to adS geometry with
$R=-2A^2$. Indeed, we have found that the constant curvature $R$ is
related to the dilaton field by $\sigma^{\prime\prime}=-R
e^{\sigma}$. Thus, with $a=1$ the equation of motion (\ref{eqmot0}) is
reduced to the much simpler equation 
\ben
\label{eqmot0a}
F(R)-RF^{\prime}(R)=0.
\een
Let us consider two more possibilities for the function $F(R)$: (i)
$F(R)=R+a R^n+b R^m+2\Lambda,$ and (ii) $F(R)=e^{-\alpha R}+2\Lambda$.
In the first case, we find that the curvature is
related to the constant parameters $a$ and $b$ via equation
$(1-n)aR^n+(1-m)bR^m+2\Lambda=0$ which shows that for $\Lambda\neq 0$
the flat space solution $R=0$ is impossible. In the second case, the
equation is $e^{-\alpha R}(1+\alpha R)+2\Lambda=0$, which admits the
flat solution $R=0$, for $2\Lambda=-1.$

We now move to four dimensional space-time. We use the same
methodology, but here the equation of motion (\ref{motgen}) takes the form 
\ben
\label{motgen1}
&&2e^{2\sigma}F(R)-
F^{\prime}(R)e^{\sigma}\frac{3}{2}\Big[2\Box\sigma+
\pa^{\alpha}\sigma\pa_{\alpha}\sigma\Big]+3\Box(e^{\sigma}F^{\prime}(R))-3\pa_{\alpha}
(e^{\sigma}F^{\prime}(R)\pa^{\alpha}\sigma)=0.
\een
The scalar curvature in this case is equal to
\ben
\label{scalcur}
R=\frac{3}{2}e^{-\sigma}
\Big[2\Box\sigma+\pa^{\alpha}\sigma\pa_{\alpha}\sigma\Big].
\een
Let us choose the function $F(R)$ again in the form
$F(R)=R+\gamma R^n+2\Lambda.$ Thus, the equation of motion takes the
form 
\ben
\label{motgen2}
&&Re^{2\sigma}+\gamma (2-n)R^ne^{2\sigma}+
3n\gamma(e^{\sigma}\Box R^{n-1}+\pa_{\alpha}R^{n-1}\pa^{\alpha} e^{\sigma})+4\Lambda e^{2\sigma}=0.
\een
We substitute the value of scalar curvature (\ref{scalcur}) to get
\ben
\label{motgen3}
&&\frac{3}{2}e^{\sigma}
\Big[2\Box\sigma+\pa^{\alpha}\sigma\pa_{\alpha}\sigma\Big]+\gamma (2-n)\left(\frac{3}{2}\right)^n e^{(2-n)\sigma}(2\Box\sigma+\pa^{\alpha}\sigma\pa_{\alpha}\sigma)^n+\\&+&2\left(\frac{3}{2}\right)^nn\gamma\left[e^{\sigma}\Box\left(
e^{-(n-1)\sigma}(2\Box\sigma+\pa^{\alpha}\sigma\pa_{\alpha}\sigma)^{n-1}\right)+\pa_{\alpha}\left(
e^{-(n-1)\sigma}(2\Box\sigma+\pa^{\alpha}\sigma\pa_{\alpha}\sigma)^{n-1}\right)\pa^{\alpha} e^{\sigma}
\right]+4\Lambda e^{2\sigma}=0.\nonumber
\een
Here we have used $\Box
e^\sigma=e^\sigma(\Box\sigma+\pa_\alpha\sigma\pa^\alpha\sigma).$ We recall that recently, in Ref.~\cite{odint} the authors
investigate cosmological solutions for $F(R)$ gravity. Here, however, we search for static, spherically symmetric solutions. Thus, we
need to replace $\Box$ by $-\Delta\simeq -{d^2}/{dr^2}-(2/r){d}/{dr}$,
and $\pa^{\alpha}A\pa_{\alpha}B$ by $-(dA/dr)(dB/dr)\equiv
-A^{\prime}B^{\prime}$. As a result, we can write 
\ben
&&2\Box\sigma+\pa^{\alpha}\sigma\pa_{\alpha}\sigma\simeq
-2\sigma^{\prime\prime}-\frac{4}{r}\sigma^{\prime}-(\sigma^{\prime})^2\equiv T.
\een
And now the equation takes the form
\ben
\label{motgen4}
&&\frac{3}{2}e^{\sigma}T+\left(\frac{3}{2}\right)^n\gamma (2-n) e^{(2-n)\sigma}T^n+
\nonumber\\&+&
2\left(\frac{3}{2}\right)^nn\gamma \Big[e^{\sigma}\left(-\frac{d^2}{dr^2}-\frac{2}{r}\frac{d}{dr}\right)\left(e^{-(n-1)\sigma}T^{n-1}\right)-\frac{d}{dr}\left(e^{-(n-1)\sigma}T^{n-1}\right)\frac{d}{dr}e^{\sigma}\Big]
+4\Lambda e^{2\sigma}=0.
\een

To solve this equation, we suppose the geometry to be adS, taking $R=-(3/2)C_0={\rm const}$. This leads us to 
\ben
\label{sigmads}
-T=2\sigma^{\prime\prime}+\frac{4}{r}\sigma^{\prime}+(\sigma^{\prime})^2=C_0e^{\sigma}. 
\een
We substitute the above equation into the equation of motion (\ref{motgen4}) to get to the simpler equation relating 
$C_0$ and $\Lambda$: $-3C_0+2\gamma (2-n)\left(-{3}C_0/2\right)^n+8\Lambda=0.$
The dilaton field being the solution of Eq.~(\ref{sigmads}) looks like
$\sigma(r)=-2\ln\,r,$ which makes $C_0$ to be zero. In this case, both
$R$ and $\Lambda$ vanish. In other words, an exact spherically
symmetric constant curvature solution is possible for $\Lambda=0$ and
corresponds to the flat space. We can go further and try another
solution of (\ref{sigmads}), for $C_0$ very small. We change
$C_0\to\epsilon,$ and consider $\sigma(r)=-2\ln(r)+\epsilon g(r).$ We
get $\sigma(r)=-2[1+(\epsilon/4)]\ln(r),$ which makes the scalar
curvature very small, but now the cosmological constant does not
vanish, being $\Lambda=(3/8)\epsilon.$ We notice that for $\epsilon$
positive, the solution leads to AdS geometry with $\Lambda$ positive,
and for $\epsilon$ negative, we get to dS geometry with $\Lambda$
negative.

Another possibility in $D=4$ is to use the {\it domain wall} ansatz in
which the dilaton depends only on a single spatial coordinate, for
instance $\sigma=\sigma(x).$ For a review on domain walls in supergravity, see Ref.~{\cite{mc}}. In this case we replace $\Box$ by
$-{d^2}/{dx^2}$, $2\Box\sigma+\pa^{\alpha}\sigma\pa_{\alpha}\sigma$ by
$-2\sigma^{\prime\prime}-(\sigma^{\prime})^2$ etc. The equation of
motion becomes 
\ben
\label{motgen3a}
&&\frac{3}{2}e^{\sigma}(-2\sigma^{\prime\prime}-(\sigma^{\prime})^2)+
\left(\frac{3}{2}\right)^n\gamma (2-n) e^{(2-n)\sigma}(-2\sigma^{\prime\prime}-(\sigma^{\prime})^2)^n+\\&+&
2\left(\frac{3}{2}\right)^nn\gamma\bigg[-e^{\sigma}\frac{d^2}{dx^2}(e^{-(n-1)\sigma}(-2\sigma^{\prime\prime}-(\sigma^{\prime})^2)^{n-1})
 -\frac{d}{dx}\left(e^{-(n-1)\sigma}(-2\sigma^{\prime\prime}-(\sigma^{\prime})^2)^{n-1}\right)e^{\sigma}\sigma^{\prime}\bigg]+4\Lambda e^{2\sigma}=0.\nonumber
\een
where prime now stands for derivative with respect to $x$. Here we used the fact that
$\frac{d}{dx}e^{\sigma}=e^{\sigma}\sigma^{\prime}$.
To solve this equation, as before we consider the ansatz $\sigma^\prime=-2Be^{\sigma/2}.$ It allows writing
\ben
\label{motflat}
2\sigma^{\prime\prime}+(\sigma^{\prime})^2=8 B^2 e^{\sigma}.
\een
This leads to $(-12B^2)^n\gamma(2-n)-12B^2+4\Lambda=0,$ and for $\Lambda\neq 0$ we can obtain the constant $B$. 

We solve the equation (\ref{motflat}) to get
\ben
\sigma(x)=-2\ln(B\,x+C),
\een
with $C$ being an integration constant. 
The scalar curvature in this case is $R=-12B^2$.

If we impose the condition that the scalar curvature (\ref{scalcur}) is constant, 
we arrive at the following reduction of the Eq.~(\ref{motgen1}): $2F(R)-RF^{\prime}(R)=0.$
Thus, if we follow the steps done in two space-time dimensions, we can
consider other possibilities for the function $F(R)$: (i) $F(R)=R+a
R^n+b R^m+2\Lambda,$ and (ii) $F(R)=e^{-\alpha R}+2\Lambda$. In
the first case, the curvature turns out to be related to the constant parameters $a$ and
$b$ via equation $R+(2-n)aR^n+(2-m)bR^m+4\Lambda=0$ which shows that for
$\Lambda\neq 0$ the flat space solution $R=0$ is impossible. In the
second case, the equation is $e^{-\alpha R}(2+\alpha R)+4\Lambda=0$,
which leads to vanishing curvature for $2\Lambda=-1.$ Note that, by
the criterion (\ref{masspert}), such solutions are always stable for
$n,m>0$. 

\section{Ending comments}
\label{sec:end}

In this work we have studied general features of $F(R)$ gravity models
and derived a formula for the asymptotic mass of such theories,
showing that for perturbative theories in which $F(R)$ allow a
perturbative expansion the value of the mass do not change
significantly from that of usual Einstein-Hilbert solutions. We
have analyzed the behavior of these theories under perturbations of
the solutions and concluded that, while metric perturbations are
essentially the same as in Einstein-Hilbert models, the scalar
curvature perturbations are unstable for positive curvature solutions
of models which modify the usual Lagrangian by $R^n$ terms with
$n<0$. It would seem that, on top of providing an effective action
mechanism by which such term would arise, one is riddled by the
stability of the solution found. Static solutions of the equations of
motion are found for the $F(R)$ gravity models restricted to the
conformal sector. We see that these solutions have the Liouville-like
kink form in two space-time dimensions. Also, very similar effects are
achieved in four-dimensional space-time, both for spherically
symmetric or domain-wall-like solutions. We also have found that the
limit $\Lambda\to0$ takes these solutions with $R\neq 0$, then still
maintaining the desired behaviour of $F(R)$ theories to present curved
backgrounds without the need for a cosmological constant. The
presence of first-order equation which solves the equation of motion
with adS geometry indicates that the system may admit supersymmetric
extensions, at least for $n$ positive. If this is indeed the case, the
supersymmetric realization will be new, given the non-standard form of
the graviton Lagrangian. This is an issue in which we will report
elsewhere.

{\bf Acknowledgements.} The authors would like to thank S.D. Odintsov
for bringing their attention to Refs.~\cite{Dolgov:2003px,Nojiri:2003ni},
and CAPES, CNPq and PRONEX/CNPq/FAPESQ for partial support. The work by AYuP is
supported by CNPq-FAPESQ DCR program, CNPq project No. 350400/2005-9.

\end{document}